\documentclass[aps,pre,twoside,centertags,showpacs,twocolumn]{revtex4}

 \usepackage{amsmath,amssymb,bm,euscript,mathrsfs}
 \usepackage{graphicx,boxedminipage}
 \usepackage[format=default,justification=RaggedRight]{caption}
 \usepackage[cp1251]{inputenc} 
 \usepackage{floatflt}
 \usepackage[hypertex]{hyperref}
 \usepackage{xcolor,fancybox}

\def\prep{Phys. Rep.\ }

\def\WRM{Waves Random Media\ }

\definecolor{lilac}{RGB}{250,0,230}
\newcommand{\bra}[1]{\ensuremath{\bm{\langle}#1\bm{|}}}
\newcommand{\ket}[1]{\ensuremath{\bm{|}#1\bm{\rangle}}}

\begin{document}

\title{The sharpness-induced mode stopping and spectrum rarefication in waveguides with periodically corrugated walls}

 \author{V.\,O. Goryashko}
 \affiliation{Uppsala University, Lagerhyddsv\"agen 1, SE-751 20, Uppsala, Sweden}
 \author{Yu.\,V. Tarasov}
 \email{yutarasov@ire.kharkov.ua}
 \author{L.\,D. Shostenko}
 \affiliation{O.\,Ya. Usikov Institute for Radiophysics and Electronics, NAS of Ukraine,\\ 12 Proskura Str., Kharkov 61085, Ukraine}

\begin{abstract}
  Starting from the rigorous excitation equation, the propagation of waves through a 2D waveguide with the periodically corrugated finite-length insert is examined in detail. The corrugation profile is chosen to obey the property that its amplitude is small as compared to the waveguide width, whereas the sharpness of the asperities is arbitrarily large. With the aid of the method of mode separation, which was developed earlier for inhomogeneous-in-bulk waveguide systems [Waves Random Media \textbf{10}, 395 (2000)], the corrugated segment of the waveguide is shown to serve as the effective scattering barrier whose width is coincident with the length of the insert and the average height is controlled by the sharpness of boundary asperities. Due to this barrier, the mode spectrum of the waveguide can be substantially rarefied and adjusted so as to reduce the number of extended modes to the value arbitrarily less than that in the absence of corrugation (up to zero), without changing considerably the waveguide average width.
\end{abstract}

\pacs{41.20.-q, 84.40.Az, 78.67.Lt, 89.90.+n}
\date{\today}
\maketitle

\section{Introduction}

The scattering through waveguides with surface roughness is a ubiquitous phenomenon which occurs on different length and time scales both in the natural environment and in different artificial systems \cite{bib:BassFuks79,bib:Maradudin07}. Whereas for environmental waveguides the problems related to wave scattering at surfaces of random nature are of major importance, for man-made guiding systems the particular attention is called by the scattering produced by the boundaries with periodical corrugation.

Periodic structures, in general, have received much attention in different areas of physics through the variety of applications of their peculiar spectra. In particular, waveguides with periodic inserts have for a long time been employed as slow-wave structures in vacuum and accelerator physics \cite{bib:Slater48,bib:LapostSept70}; photonic (Bragg) crystals are widely used as frequency selective narrowband resonators and filters in terahertz and optical ranges \cite{bib:TecHolldElias10,bib:BarillDiligBenedMerlo06,bib:Bykov72}; multilayered films with periodic variation of the refractive index serve as effective mirrors in the VUV and X ranges of EM radiation \cite{bib:XrayOpt,bib:MichWiesm_etal01}. The importance of periodic structures for numerous applications and the diversity of physical effects originating in them have stimulated the development of a number of theoretical methods for their analysis. The detailed review of the methods developed up to the mid fifties of the 20-th century can be found in Ref.~\cite{bib:Brilluin53}.

For periodic structures, apart from the foremost question of how the forbidden gaps in their spectra originate, there arises a number of other intriguing questions regarding their physical properties, especially for those systems whose inhomogeneities are much less in size than the operating wavelength. Until recently it was believed that inhomogeneities of this sort cannot crucially affect the wave propagation. For this reason the studies on structures with sub-wavelength periodic variations of their parameters have for long remained beyond the main line of research until the previous decade, when anomalous wave transmission through sub-wavelength hole arrays in optically thick films was detected \cite{bib:Ebbesen98,bib:Ebbesen10}. In our recent paper \cite{bib:GanTarShost11} it was also discovered that the small-scale distortion of wave resonator boundaries dramatically affects their spectra. Specifically, it was found that modal content of rough-side resonators was mainly controlled by the sharpness of boundary asperities rather than by their height. Even for the asperities of sub-wavelength nature the spectrum may be adjusted in such a way that at any desirable frequency range the quasioptical resonator can become even a single-frequency one.

The results of Ref.~\cite{bib:GanTarShost11} were obtained for closed systems only, being related to their spectral properties. It would be, however, of great interest to find out if small boundary asperities could produce similar radical effects on the properties of \emph{open} systems, in particular, on the wave transmission in waveguide structures. Yet, in trying to answer this question one faces the problem which has not so far been adequately resolved. Specifically, any self-consistent theory of a waveguide system with arbitrarily corrugated side walls is relied unavoidably on the results related to plane-wave scattering by such boundaries. For boundaries whose corrugation is random such theories do exist for a long time. Yet they are all known to allow obtaining the tractable and practically useful results only in the case of smooth roughness, i.\,e., for the asperities whose mean-square tangent of slope is small as compared with unity \cite{bib:BassFuks79,bib:Simonsen10}. For cavity resonators, this restriction was overcome in Ref.~\cite{bib:GanTarShost11}, but for systems of general type, including waveguides and different surfaces subject to radiation, this is yet to be done.

In studying the wave scattering by distorted boundaries the most challenging task is, in general, related to boundaries' random character. The scattering by the surfaces whose corrugation is  deterministic, in particular, periodic in tangent coordinate seems to be, at first glance, a less intriguing problem. Normally, the periodicity as such is considered to be the main feature of such systems, whereas the the specific details like the amplitude and the sharpness of corrugation receive less attention. By now, there has been a number of works on wave propagation in periodically corrugated waveguides (see, e.\,g., Refs.~\cite{bib:WuSprungMart93,bib:Pogrb98,bib:Pogrb04prb,bib:Pogrb04,bib:OkitaTanaka12,bib:BarraPagneuxZuniga12} and references therein). Many peculiarities of their spectra are studied in more detail. But in all of the aforesaid papers the particular effective-medium models were used.  In spite of the fact that these models allow to take adequately into account the periodicity property, they were not capable of allowing for other substantial features of the wave scattering by uneven surfaces.

Meanwhile, in Refs.~\cite{bib:MakTar98,bib:MakTar01} devoted to charge transfer through randomly rough single-mode quantum waveguides it was revealed that the particular role in the transport and spectral properties of such systems was played by the so-called ``slope'', or ``gradient'', scattering mechanism, in which as the dominant factor for surface-roughness-induced scattering of propagating waves serves the degree of asperity smoothness. The same scattering mechanism was found to influence radically the spectra of cavity resonators with sharply rough walls~\cite{bib:GanTarShost11}. As far as the principal difference between random and deterministic corrugation lies in the difference in correlation lengths, the results of Refs.~\cite{bib:MakTar98,bib:MakTar01,bib:GanTarShost11} suggest quite strong grounds to believe that the gradients of boundary roughness will have a great impact on the properties of periodically corrugated waveguides as well.

To verify this conjecture, in this study we solve the problem of wave transport through the waveguide with periodically corrugated finite-length segment whose corrugation is thought of as being small in amplitude but arbitrarily large in sharpness. Through the application of special operator technique of transverse-mode separation the transport of each of the transverse mode in the waveguide with corrugated insert is shown to be equivalent to quantum-particle transport through the individual sharply modulated potential barrier whose average magnitude is governed by the degree of asperity sharpness. The small-scale nature of the asperities makes it possible to operate in the lowest Bloch band only, which allows to concentrate the attention on the sharpness-induced scattering leaving the band-forming Bragg scattering beyond consideration. Within the suggested approach it is revealed that the slight-in-amplitude modulation of the waveguide boundaries, taken to be not extremely smooth, affects significantly the waveguide mode structure.  On sharpening the asperities the number of propagating modes reduces substantially, the fast modes turning sequentially, first, into the slow and then into evanescent modes. In such a way the multimode waveguide can be made single-mode or even evanescent-mode one without noticeable variation of its local width.

\section{The problem formulation}

\begin{figure}[h]
  \setcaptionmargin{.3in}%
  \centering
  \scalebox{.7}[.7]{\includegraphics{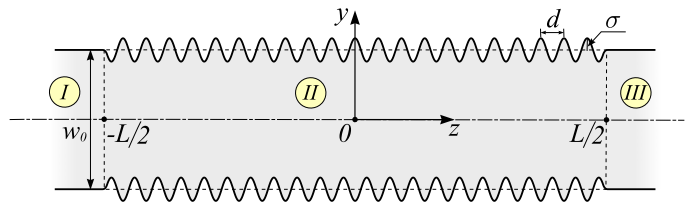}}
 \caption{(Color online) Two-dimensional waveguide with periodically corrugated insert (segment ${z\in[-L/2,L/2]}$).
  \hfill\label{fig1}}
\end{figure}

We will consider the infinitely long two-dimensional waveguide with straight parallel boundaries, whose side walls at some finite segment of length $L$ are subjected to periodical corrugation (see Fig.~\ref{fig1}). The width of the entire waveguide will be specified by the function $w(z)=w_0+2\xi(z)\theta(L/2-|z|)$,
where $w_0$ is the width of the straight segments \textit{I} and \textit{III}, $\xi(z)$ is the periodic function with period $d$, amplitude $\sigma$ and zero mean value. To simplify further numerical calculations, but without substantial loss of generality, we will describe the corrugation by function $\xi(z)=\sigma\cos(2\pi z/d)$, taking the amplitude small as compared to waveguide average width,
\begin{equation}\label{Small_ampl}
  \sigma\ll w_0\ .
\end{equation}

Consider the experimental setup when in a real 3D waveguide the TE type waves are excited, namely, the TE$_{0n}$ modes with indices corresponding to field variations along $x$ and $y$ axes, respectively. In this case, all the electromagnetic field components can be found with the aid of one ($x$) component of vector potential $\mathbf{A}(y,z)$ that obeys the homogeneous wave equation. The waveguide side boundaries will be considered as ideally conducting, which for $s$-polarized waves corresponds to Dirichlet boundary conditions at surfaces $y=\pm w(z)/2$  while on the straight $x$-coordinate boundaries (not shown in Fig.~\ref{fig1}), if they are taken into consideration, the Newmann conditions should be met. As far as the conditions are concerned at the open waveguide ends, we will discuss them below, after the main equation for the sought-for field will be written down in the mode representation.

To solve the excitation problem for the waveguide we will seek for the Green function which obeys the following equation
\begin{equation}\label{Helmholtz_Green}
  (\Delta+k^2)G(\mathbf{r},\mathbf{r}')=\delta(\mathbf{r}-\mathbf{r}')\ ,
\end{equation}
where $\Delta$ is the two-dimensional Laplace operator. As a first step, we Fourier transform equation \eqref{Helmholtz_Green} with respect to the locally (in the coordinate $z$) complete set of eigenfunctions of the transverse Laplacian, viz.
\begin{equation}\label{Whole_set}
  \ket{y,z;n} =\left[\frac{2}{w(z)}\right]^{1/2}
  \sin\left\{\left[\frac{y}{w(z)}+\frac{1}{2}\right]\pi n\right\}\ ,
  \qquad n\in\aleph\ ,
\end{equation}
thus expressing the Green function as a double series
\begin{equation}\label{Green_ModeRep}
G(\mathbf{r},\mathbf{r}') = \sum_{n,n'=1}^{\infty}
\bra{y,z;n}G_{nn'}(z,z')\ket{y',z';n'} \ .
\end{equation}
For mode coefficients $G_{nn'}(z,z')$ the following set of coupled equations results from Eq.~\eqref{Helmholtz_Green},
\begin{widetext}
\begin{equation}\label{G_mode-eqn}
  \left[\frac{\partial^2}{\partial z^2}+
  k^2_n-V_n(z)\right]G_{nn'}(z,z')
  -\sum_{m=1\atop (m\neq n)}^{\infty}\hat{U}_{nm}(z)
  G_{mn'}(z,z')=
  \delta_{nn'}\delta(z-z') \ ,
\end{equation}
\end{widetext}
where $k_n^2=k^2-\left(\pi n/w_0\right)^2$ is the longitudinal propagation number for the $n$-th waveguide mode in the non-corrugated outward parts of the waveguide, $V_n(z)$ and $\hat{U}_{nm}(z)$ are the effective ``potentials'' (in general, of operator nature) which arise in the wave equation due to the boundary corrugation,
\begin{subequations}\label{Potentials_VnUnm}
 \begin{align} \label{Vn}
  & V_n(z)=\pi^2n^2\left[\frac{1}{w^2(z)}-\frac{1}{w_0^2}\right]+
   \left(1+\frac{\pi^2n^2}{3}\right)
   \left[\frac{w'(z)}{2w(z)}\right]^2,\\
  & \hat{U}_{nm}(z)=2B_{nm}\Bigg\{D_{nm}\left[\frac{w'(z)}{w(z)}\right]^2\notag\\
  \label{Unm}
  &\qquad\qquad\qquad\quad  -   \frac{1}{w(z)}\left[w'(z)\frac{\partial}{\partial z}+\frac{\partial}{\partial z}w'(z)\right]\Bigg\}\,.
 \end{align}
\end{subequations}
These potentials are concentrated across the finite interval, so we will regard them as the sources for intra- and intermode scattering of outward waveguide modes, respectively. The numerical coefficients in potential \eqref{Unm} have the form
\begin{equation}\label{BnmDnm}
 \begin{split}
   &B_{nm}=\frac{nm}{n^2-m^2}\cos^2\left[\frac{\pi}{2}(n-m)\right]\ ,\\
   &D_{nm}=\frac{3n^2+m^2}{n^2-m^2}\ .
 \end{split}
\end{equation}

Equation \eqref{G_mode-eqn} must be supplemented with boundary conditions (BC)'s along axis $z$. Since the waveguide is an open-ended system, as the BC's at $z\to\pm\infty$ we will use the radiation conditions. To formulate them analytically for a non-homogeneous waveguide system is a challenging task in the general case. However, in view of the perturbation potentials in Eq.~\eqref{G_mode-eqn} being identically equal to zero in the regions $|z|>L/2$, the Green function mode structure in these regions reduces to its diagonal components. So, the legible boundary conditions may be \textit{a~priori} formulated for these components only. The next section will show that this is sufficient for determining the entire Green function since all the non-diagonal components of its mode matrix can be lineally expressed in terms of diagonal elements, for which we obtain the whole set of \emph{uncoupled} equations.

\section{Mode separation in the waveguide with nonuniform segment}
\label{ModeSep}

The set of equations \eqref{G_mode-eqn} may be solved for mode components of the Green function using the special operator technique applicable for potentials $V_n(z)$ and $\hat{U}_{nm}(z)$ of quite an arbitrary form. The technique was first elaborated in Ref.~\cite{bib:Tar00} for the open systems of waveguide configuration and then expanded to open-ended and closed resonator-type systems disordered in the bulk \cite{bib:GanErTar07} as well as on the surface \cite{bib:GanErTar09,bib:GanTarShost11}. With regard to the waveguide-type systems, the method reduces (schematically) to the following action sequence. At the first stage, by putting $n\neq n'$ in the set of equations \eqref{G_mode-eqn} one should resolve this set with respect to the off-diagonal components of mode matrix $\|G_{mn}\|$, thereby expressing them, by means of linear operation, through the diagonal elements whose mode indices are coincident with the right-hand (column) index of the required off-diagonal element,
\begin{equation}\label{G_mn->G_nn}
  G_{mn}(z,z')=\int_L\mathrm{d}z_1\,
  \mathsf{K}_{mn}(z,z_1)G_{nn}(z_1,z')
  \qquad(m\neq n) \ .
\end{equation}
In Eq.~\eqref{G_mn->G_nn}, the integration runs over the axis $z$ region where potentials \eqref{Potentials_VnUnm} are not equal to zero. The kernel of the integral operator in~\eqref{G_mn->G_nn} can be found from the equation of Lippmann-Schwinger type,
\begin{equation}\label{K_numu}
  \mathsf K_{mn}(z,z')={\mathsf R}_{mn}(z,z')
  +\sum_{k\neq n}
  \int_Ldz_1\,\mathsf{R}_{mk}(z,z_1)
  \mathsf{K}_{kn}(z_1,z') \ ,
\end{equation}
where we introduce the notation
\begin{equation}\label{kernR}
  \mathsf{R}_{mn}(z,z')=G^{(V)}_{m}(z,z')U_{mn}(z')
\end{equation}
for matrix elements of some operator $\hat{\mathsf{R}}$ whose domain of definition is two-dimensional space $\mathsf{M}=\{z,m\}$ which includes the coordinate axis ($z$) and the whole set of mode indices ($m$). This operator apparently accounts for scattering between all the different transverse modes, so in what follows we will refer to it as the mode-mixing operator.

Function $G^{(V)}_{m}(z,z')$ in expression \eqref{kernR}, which we identify as the \emph{trial} mode Green function, is the solution of equation
\begin{equation}\label{G(V)_m-eq}
    \left[\frac{\partial^2}{\partial z^2}+
  k^2_n-V_n(z)\right]G^{(V)}_n(z,z') =\delta(z-z')\ ,
\end{equation}
which differs from Eq.~\eqref{G_mode-eqn} by the lack of intermode potentials (to unify the notations in the subsequent formulas we re-designate the mode variable of the Green function in Eq.~\eqref{G(V)_m-eq} in comparison with Eq.~\eqref{kernR}, viz. $m\to n$).

Upon solving equation \eqref{G(V)_m-eq} and considering the operator $\hat{\mathsf{R}}$ matrix elements as already known functions, one can substitute off-diagonal elements of the Green matrix into equation \eqref{G_mode-eqn}, thereby obtaining closed equations for the diagonal elements of this matrix (for $\forall n$'s),
\begin{equation}\label{GDIAG-FIN}
  \left[\frac{\partial^2}{\partial z^2}+
  k^2_{n}-V_{n}(z)-\hat{\mathcal{T}}_{n}\right]
  G_{nn}(z,z')=\delta(z-z') \ .
\end{equation}
In equation \eqref{GDIAG-FIN}, if compared with Eq.~\eqref{G(V)_m-eq} for the trial Green function, the additional \emph{operator} potential $\hat{\mathcal{T}}_{n}$ has appeared, whose expression has the form
\begin{equation}\label{T-oper}
  \hat{\mathcal{T}}_{n}=\bm{P}_{n}\hat{\mathcal U}
  (\openone-\hat{\mathsf R})^{-1}\hat{\mathsf R}\bm{P}_{n}\ .
\end{equation}
Here, $\hat{\mathcal U}$ is the intermode operator potential which is specified in $\mathsf{M}$ by matrix elements
\begin{equation}\label{U-matr}
  \bra{z,n}\hat{\mathcal U}\ket{z',n'} =
  {U}_{nn'}(z)\delta(z-z')\ ,
\end{equation}
$\bm{P}_{n}$ is the projection operator, whose action reduces to the assignment of given value $n$ to the nearest mode index of any operator standing next to it, no matter if it is to the left or to the right. Two embracing projectors in Eq.~\eqref{T-oper} have appeared due to the restriction of summation in Eqs.~\eqref{K_numu} and \eqref{G_mode-eqn} by mode indices $k\neq n$. The operators standing between the projectors act in the subspace ${\mathsf{\overline{M}}_{n}}\in\mathsf{M}$ which includes the coordinate axis $z$ and the set of all mode indices other than the separate index $n$. The role of the projectors in operator potential Eq.~\eqref{T-oper} is to reduce the action of this operator on the function standing to the right of it solely to integration over~$z$, without summation over mode indices.

From the functional structure of potential \eqref{T-oper} one can infer that in equation \eqref{GDIAG-FIN} for the diagonal propagator of a given transverse mode this potential accounts actually for the \emph{intermode} scattering. In contrast to potential $V_n(z)$, the potential $\hat{\mathcal{T}}_{n}$, in the general case, acts as the non-local (in $z$) operator whose characteristic scale is determined by the spatial extent of the trial Green functions and the intermode potentials, both entering the $T$-potential through the mode-mixing operator.

As regards the boundary conditions to equation \eqref{GDIAG-FIN}, the $n$-th waveguide mode, whose dynamics is governed by this equation, is proven to be effectively separated from other modes, which are present in Eq.~\eqref{GDIAG-FIN} only implicitly, as the intermediate scattering modes ``hidden'' inside the $T$-potential. If Eq.~\eqref{GDIAG-FIN} is recast in the form of the integral Dyson equation, where $G^{(V)}_n(z,z')$ stands as the unperturbed sought-for solution, it will be clear that open boundary conditions for the exact intramode Green function coincide with those satisfied by the corresponding trial function. The conditions for the latter function will be presented in the next section.

Thus, taking account of the relationship \eqref{G_mn->G_nn} between diagonal and off-diagonal elements of the mode Green matrix, which can be recast in the operator form
\begin{equation}\label{Gmn->Gnn}
  \hat{G}_{mn}=\bm{P}_m(\openone-\hat{\mathsf R})^{-1}\hat{\mathsf R}\bm{P}_n \hat{G}_{nn}\ ,
\end{equation}
with $\hat{G}_{mn}$ and $\hat{G}_{nn}$ being the operators in the coordinate variable $z$, one can assert that the solution of the entire set of uncoupled equations \eqref{GDIAG-FIN} (for $\forall n$'s) completely determines the sought-for Green function of the waveguide with the inhomogeneous insert.

Upon deriving equation \eqref{GDIAG-FIN} and explicit expressions for the potentials entering it, we in fact have performed the effective ``mode separation'' in the original two-dimensional Helmholtz equation. This operation substantially facilitates the solution of the problem of calculating the fields in the originally two-dimensional inhomogeneous waveguide since there exists a number of advanced mathematical methods to solve wave equations in the dimension one.

However, it should be admitted that at the expense of the achieved simplification, in the master equation, whose role from now on will be played by equation \eqref{GDIAG-FIN}, we have obtained the additional (to the original $V_n(z)$) intramode potential with quite non-trivial functional structure. The effective potential of the analogous type is familiar in the quantum scattering theory \cite{bib:T75,bib:N68}, being referred to as \emph{T}-matrix. However, the conventionally defined \emph{T}-matrix is known to be an extremely singular mathematical object, whose calculation is normally carried out in the lowest order of perturbation theory. In our calculation technique the $T$-potential in Eq.~\eqref{GDIAG-FIN} has no singularities. In Ref.~\cite{bib:Tar00} it was shown that if the effective potentials in Eq.~\eqref{G_mode-eqn} are random functions, which break the spatial symmetry of the system under consideration, the regularity of potential \eqref{T-oper} is ensured by splitting the entire set of effective potentials into the subsets of intra- and intermode ones. But the system we consider in this study is subject to regular rather than random perturbations, so it is not \emph{a~priori} clear if the arguments of Ref.~\cite{bib:Tar00} are appropriate for this case.

At the same time, the system we deal with is the open system and thus non-Hermitian, which normally is sufficient for the inverse operator in Eq.~\eqref{T-oper} to be nonsingular. The proof of this fact can be found in fundamental papers by Feshbach on nuclear reaction theory (see also Refs.~\cite{bib:Feinberg09,bib:Feinberg11}, where this issue is discussed in detail with regard to one-dimensional systems subject to final-support external potentials). Yet, to additionally protect oneself against the degeneracy property of the full wave operator in Eq.~\eqref{G_mode-eqn} one can always supply the ``energy'' $k^2$ with additional infinitesimal positive-imaginary term which accounts for the dissipation and certainly ensures the lack of degeneracy, thus rendering the inverse operator in Eq.~\eqref{T-oper} properly defined.

\section{Calculation of the intramode propagators}
\label{Green_func_calc}

As far as the sought-for Green function \eqref{Green_ModeRep} is completely determined by the diagonal components of its mode matrix, from here on we will regard the solution of Eq.~\eqref{GDIAG-FIN} as the main object of our study. Under fixed mode index $n$ this equation can be thought of as being uncoupled provided that all trial Green functions $G^{(V)}_m(z,z')$ with mode indices $m\neq n$ are beforehand determined. The next section is devoted just to the solution of this auxiliary problem, namely, to finding the trial functions.

\subsection{The trial Green function}

Function $G^{(V)}_n(z,z')$ solves one-dimensional equation \eqref{G(V)_m-eq} on the condition that both ends of the waveguide are open. This condition, referred to in electrodynamics as Sommerfeld's radiation condition \cite{bib:BassFuks79,bib:Vladimirov67}, can be formulated in the form of two equalities connecting the sought-for function and its first derivative, viz.
\begin{equation}\label{BC_trial_GF}
  \left.\left(\frac{d}{dz}\mp ik_n\right)G^{(V)}_n(z,z')\right|_{z\to\pm\infty}=0\ .
\end{equation}
Being so posed, the problem of finding the trial Green function pertains to the class of \emph{boundary-value} problems of open type, whose solution is difficult to obtain even for potentials of relatively simple form.

Fortunately, in the case of one spatial dimension the boundary-value Green function problem reduces to the pair of problems of causal type (see, e.\,g., Ref.~\cite{bib:MorseFeshbach53}). This can be done through representation
\begin{align}
  G_n^{(V)}(z,z')= & \left[\mathcal{W}_n(z')\right]^{-1}
  \big[ \psi_+(z|n)\psi_-(z'|n)\Theta(z-z') \notag\\
  & +\psi_+(z'|n)\psi_-(z|n)\Theta(z'-z) \big]\ ,\label{Green-Cochi}
\end{align}
where functions $\psi_{\pm}(z|n)$ are the solutions of two auxiliary Cauchy problems for homogeneous equation \eqref{G(V)_m-eq}, whose boundary conditions are specified on solely one (``plus'' or ``minus'', respectively) end of the interval where Green function is being sought for; $\mathcal{W}_n(z)$ is the Wronskian of those solutions, which, in concordance with Liouville's  formula \cite{bib:Pontr74}, for equation \eqref{G(V)_m-eq} with potential \eqref{Vn} does not depend on the coordinate $z$.

In this paper we choose the potential $V_n(z)$ to be a periodic function in the range ${z\in[-L/2,L/2]}$. Basically, this enables one to explore the solution of Eq.~\eqref{G(V)_m-eq} taking advantage of Floquet theory \cite{bib:Floquet1883}. Yet, the structure of the right-hand side (r.h.s.) of Eq.~\eqref{Vn} is excessively complicated for this purpose, so we will make some simplifications allowing for the reduction of Eq.~\eqref{G(V)_m-eq} to the standard Mathieu equation.

\begin{figure}[h!]
  \centering
  \scalebox{.4}{\includegraphics{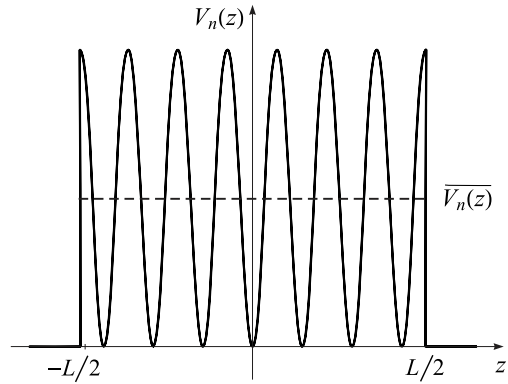}}
  \caption{The effective potential barrier in equation \eqref{G(V)_m-eq} arisen due to the waveguide corrugation over the finite interval $L$.
  \label{fig2} }
\end{figure}

We will first regard geometric parameters of corrugation as obeying the inequalities
\begin{equation}\label{Gofr_parameters}
  \sigma,\,d\ll w_0\ .
\end{equation}
At the same time, the sharpness of corrugation will be considered as arbitrary. By separating in the potential \eqref{Vn} its mean value and the term that oscillates about zero, $V_n(z)=\overline{V_n(z)}+\Delta V_n(z)$, under conditions \eqref{Gofr_parameters} we have
\begin{subequations}\label{V_n=two_terms}
\begin{align}\label{V_n-average}
  & \overline{V_n(z)}=
  \frac{2\pi^2}{w_0^2}\left(1+\frac{\pi^2n^2}{3}\right)\Xi^2\ ,\\
\label{V_n-oscillate}
  & \Delta V_n(z) = - \frac{2\pi^2}{w_0^2}\left(1+\frac{\pi^2n^2}{3}\right)\Xi^2 \cos\frac{4\pi z}{d}\ ,
\end{align}
\end{subequations}
where parameter $\Xi=\sigma/d$ is introduced, which specifies the degree of the asperity sharpness. The sum of potentials \eqref{V_n=two_terms} is shown schematically in Fig.~\ref{fig2}, making us to expect that the corrugation of the finite waveguide portion would result in the effects substantially analogous to those arising in the quantum particle transport through periodically
modulated potential barrier. The particular feature of our problem is that the average height of the effective potential barrier, being proportional to the square of parameter $\Xi$, appears to be governed by the sharpness of the corrugation profile. Below we give a brief account of the calculation procedure for causal functions $\psi_{\pm}(z|n)$ and, correspondingly, for Green function \eqref{Green-Cochi}.

Consider, for definiteness, function $\psi_{+}(z|n)$; function $\psi_{-}(z|n)$ can be found analogously. By going over to the dimensionless coordinate $t=2 \pi z/d$ the equation for the sought-for function reduces to the canonical Mathieu equation~\cite{bib:AbramStegun64},
\begin{equation}\label{Mathieu_equation}
   \psi_+''(t|n) + \big[a_n - 2 q_n \cos (2t)\big] \psi_+(t|n) =0 \ .
\end{equation}
Here, prime signs stand for the derivatives over $t$-variable and the notations are used
\begin{subequations}\label{a_n,q_n}
\begin{align}
  a_n & =\frac{d^2}{4 \pi^2}\left[k^2_n -\overline{V_n(t)}\right] =
  \frac{d^2}{4 \pi^2} k_n^2 + 2q_n\ ,\\[6pt]
  q_n & = -\left(1+\frac{\pi^2n^2}{3}\right)\left(\frac{\sigma}{2 w_0}\right)^2\ .
\end{align}
\end{subequations}

Bearing in mind that at the ``plus'' end of the waveguide function $\psi_{+}(z|n)$ obeys radiation condition similar to Eq.~\eqref{BC_trial_GF} we will seek this function in the form
\begin{subequations}\label{psi+Bloch_sol}
\begin{equation}\label{psi+}
\psi_+(z|n)=\left\{
 \begin{aligned}
  & \mathrm{e}^{i k_n (z+L/2)} + r_n \mathrm{e}^{-i k_n (z+L/2)}\ , \!\!\!\! &&\ z  \leq -L/2\\
  & A_n\varphi_n(z) +  B_n \varphi_n(-z)\ , && |z| < L/2\ \ ,\\
  & t_n \mathrm{e}^{i k_n (z-L/2)}\ , &&\ \ z \geq  L/2
 \end{aligned}\right.\ 
\end{equation}
where
\begin{equation}\label{Bloch_sol}
   \varphi_n(z)=\mathrm{e}^{i\varkappa_nz}g(z|n)
\end{equation}
\end{subequations}
is the standard Floquet-Bloch solution of equation \eqref{Mathieu_equation} on the entire (infinite) axis $z$; $A_n$, $B_n$, $r_n$ and $t_n$ are the constants subject to determination from the matching conditions at the junction points between corrugated and straight waveguide portions. Bloch wavenumber ${\varkappa_n=2 \pi \mu_n/d}$ in the exponential in Eq.~\eqref{Bloch_sol} is related to characteristic number $\mu_n$ of equation \eqref{Mathieu_equation}, being the function of parameters $a_n$ and $q_n$~\cite{bib:AbramStegun64}. Functions $\varphi_n(\pm z)$ in the second line of Eq.~\eqref{psi+} are the linearly independent solutions of the wave equation on the interval of waveguide corrugation. Function $g(z|n)$ from Eq.~\eqref{Bloch_sol} is, in the general case, representable as a series,
\begin{equation}\label{g(z|n)}
  g(z|n) = \frac{1}{f_0^{(n)}}\sum_{m=-\infty}^\infty f_m^{(n)} \mathrm{e}^{ 4 \pi i m z/d}\ ,
\end{equation}
whose coefficients are specified by the parameters of equation \eqref{Mathieu_equation} and obey the following set of recurrence relations,
\begin{align}
 & \left[a_n - (\mu_n + 2 m)^2 \right]  f_m^{(n)}
  + {q_n} f_{m-1}^{(n)}
  + {q_n} f_{m+1}^{(n)} = 0 \\
 &\ (m= \ 0,\ \pm 1,\ \pm 2,\ \dots)\ .\notag
\end{align}

\begin{figure}[h!]
  \centering
  \scalebox{.65}{\includegraphics{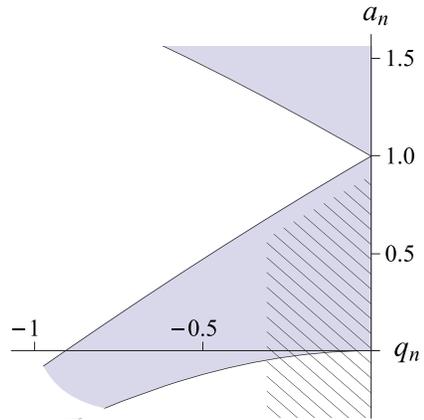}}
  \caption{Stability areas (grey) for equation Eq.~\eqref{Mathieu_equation} solutions. The shading approximately covers the region corresponding to inequalities \eqref{q_n<<1_a_n<1}.
  \label{fig3} }
\end{figure}

In the present study, we focus our attention on the effect produced on the spectral and dynamical properties of the waveguide by the gradient scattering mechanism \emph{per~se}, without making an emphasis on spectrum multi-band structure associated with the periodic nature of waveguide corrugation. Therefore, in what follows we will restrict ourselves by the values of parameters $a_n$ and $q_n$ that fall into the shaded region in Fig.~\ref{fig3} corresponding to the couple of inequalities,
\begin{equation}\label{q_n<<1_a_n<1}
  |q_n|\ll 1\ ,\qquad\qquad a_n < 1 - |q_n|\ .
\end{equation}
The region includes the extended states of only first stability band of solutions to equation
\eqref{Mathieu_equation}, as well as the evanescent states with negative mode energies. In terms of physical parameters of the system at hand, the inequalities \eqref{q_n<<1_a_n<1} suggest the relative smallness of the corrugation amplitude and the smallness of its period against the wave length, i.\,e., the fulfillment of inequalities
\begin{equation}\label{sigma<<w_0/Nc}
  \sigma\ll\frac{w_0}{N_c}\ ,\qquad\lambda=\frac{2 \pi}{k}> d\ ,
\end{equation}
where $N_c$ is the number of extended modes in the corrugated region. Under conditions \eqref{sigma<<w_0/Nc}, the terms of series \eqref{g(z|n)} decrease rapidly with growing
$|m|$, thus permitting us to restrict the summation by three terms only, with indices ${m=0,\pm 1}$. In this case the characteristic number of equation \eqref{Mathieu_equation} differs insignificantly from its value at $q_n=0$, so in the estimations one can always put $\mu_n \approx \sqrt{a_n}$.

The number of extended trial modes in the corrugated segment of the waveguide is determined by the condition of positiveness of the entire mode ``energy'', $\varkappa_n^2\approx k_n^2 - \overline{V_n(z)}$, and therefore is largely dependent on the corrugation profile sharpness,
\begin{equation}\label{N_c0}
  N_c(\Xi)=\left[\sqrt{\frac{N_{c\,0}^2-2\Xi^2}{1+(2\pi^2/3)\Xi^2}}\right]\ .
\end{equation}
In Eq.~\eqref{N_c0}, square brackets denote an integer part of the enclosed number, $N_{c\,0}=[kw_0/\pi]$ is the number of extended modes in the corrugation-free waveguide parts.
\begin{figure}[h!]
  \centering
  \scalebox{.8}{\includegraphics{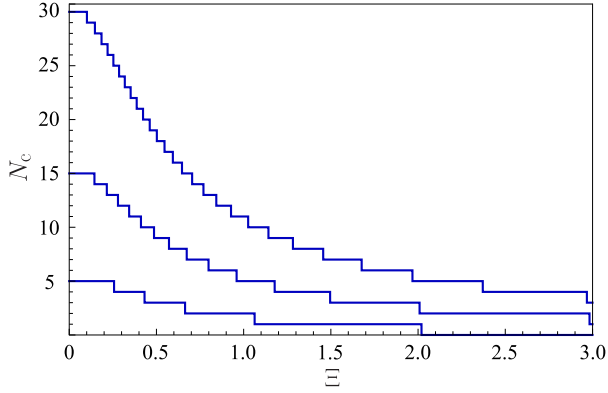}\qquad}
  \caption{The number of extended modes in the corrugated waveguide segment versus the sharpness of corrugation profile. The diagrams correspond to  $N_{c\,0}=5,15,\ \text{and}\ 30$.
\label{fig4} }
\end{figure}
In Fig.~\ref{fig4}, the diagrams of $N_c(\Xi)$ are presented for several specific values of the average waveguide width. It is easily seen that on sharpening the asperity profile the number of extended modes in the corrugated region decreases quite rapidly, reaching zero at some critical value of the sharpness parameter $\Xi$, namely, at $\Xi_{cr}=N_{c\,0}/\sqrt{2}$. The wavenumber $\varkappa_n$ in Eq.~\eqref{Bloch_sol} is real-valued at $\Xi<\Xi_{cr}$, while at $\Xi>\Xi_{cr}$ the corrugated segment of the waveguide becomes practically ``non-transparent'', in the sense that all modes within it turn into the evanescent ones, each decreasing exponentially over the scale of the order of $|\varkappa_n|^{-1}$.

By moving along axis $z$ in Fig.~\ref{fig1} from region III in the negative direction and matching the fields sequentially at boundaries between regions III--II and II--I we obtain the set of equations for coefficients entering Eq.~\eqref{psi+}. The solution of this set reads,
\begin{subequations}\label{t_n+r_n+_0}
  \begin{alignat}{2}\label{t_n}
   & t_{n} = \frac{\left(1-\mathcal{R}_n^2\right)
    \exp\big(i\varkappa_n L\big)}
    {1-\mathcal{R}_n^2
    \exp\big(2i\varkappa_n L\big)}
    \ ,\\
\displaybreak
  \label{r_n}
   & r_{n} =\mathcal{R}_n
    \frac{1-\exp\big(2i\varkappa_nL\big)}
    {1-\mathcal{R}_n^2
    \exp\big(2i\varkappa_n L\big)}\ ,\\
  \label{A_n}
   & A_n =t_n \frac{\beta_n +   k_n}
    {2 g_n \beta_n}  \mathrm{e}^{-i \varkappa_n L/2}\ ,\\
  \label{B_n}
   & B_n =t_n \frac{\beta_n -  k_n}
    {2 g_n \beta_n} \mathrm{e}^{i \varkappa_n L/2}\ .
  \end{alignat}
\end{subequations}
In Eqs.~\eqref{A_n} and \eqref{B_n} we have introduced the notations
\begin{equation}\label{beta_n;g_n}
\begin{aligned}
 & \beta_n =  \varkappa_n - i  \frac{g'(L/2|n)}{g(L/2|n)}\ \\
 & g_n=g(L/2|n)\ .
\end{aligned}
\end{equation}
Note that in deriving Eqs.~\eqref{t_n+r_n+_0} the waveguide side boundaries were assumed to be continuous at points $z=\pm L/2$, but not necessarily smooth. It can be easily verified that coefficients $t_n$ and $r_n$, Eqs.~\eqref{t_n} and \eqref{r_n}, relate with one another via equality ${|t_n|^2+|r_n|^2=1}$, which suggests the validity of their interpretation in Eq.~\eqref{psi+} as the mode transmission and reflection coefficients, respectively.

The calculation scheme for function $\psi_{-}(z|n)$ is completely analogous to that for $\psi_{+}(z|n)$, differing only by the inversion of axis $z$. By performing the calculations and by inserting the obtained expressions for $\psi_{\pm}(z|n)$ into formula~\eqref{Green-Cochi}, the trial Green function in the corrugated waveguide segment can be expressed as a sum of four terms each of which is proportional to the product of Bloch eigenfunctions running in opposite directions, viz.
\begin{widetext}
\begin{align}\label{TrialGreenFunction_2}
 G^{(V)}_n (z,z') &=
 \mathcal{G}_{11}^{(n)} \mathrm{e}^{i \varkappa_n (z-z')} g(z|n)g(-z'|n) +
 \mathcal{G}_{22}^{(n)} \mathrm{e}^{-i \varkappa_n (z-z')} g(-z|n)g(z'|n) \nonumber\\
 &\quad + \mathcal{G}_{12}^{(n)} \mathrm{e}^{i \varkappa_n (z+z')} g(z|n)g(z'|n) +
 \mathcal{G}_{21}^{(n)} \mathrm{e}^{-i \varkappa_n (z+z')} g(-z|n)g(-z'|n)\ .
\end{align}
\end{widetext}
Factors $\mathcal{G}_{ik}^{(n)}$ in Eq.~\eqref{TrialGreenFunction_2} have the form
\begin{subequations}\label{G_ik}
\begin{alignat}{2}
  \label{G_11}
  \mathcal{G}_{11}^{(n)} &=- i\frac{\mathcal{Q}_n }{2 g_n^2 \beta_n}\Big[\theta(z-z') + \mathcal{R}_n^2 \mathrm{e}^{2 i\varkappa_n L} \theta(z'-z)\Big]\ ,\\
  \label{G_22}
  \mathcal{G}_{22}^{(n)} &=-i\frac{\mathcal{Q}_n }{2g_n^2 \beta_n}\Big[\theta(z'-z) + \mathcal{R}_n^2 \mathrm{e}^{2 i\varkappa_n L} \theta(z-z')\Big]\ ,\\
  \label{G_121}
  \mathcal{G}_{12}^{(n)} &=
  \mathcal{G}_{21}^{(n)}= i\frac{\mathcal{Q}_n }{2g_n^2 \beta_n}\mathcal{R}_n\mathrm{e}^{i\varkappa_n L}\ ,
\end{alignat}
\end{subequations}
and the notations are used
\begin{subequations}
\begin{align}
 \label{R_n}
 & \mathcal{R}_n = \frac{k_n -\beta_n}{k_n +\beta_n}\ ,\\
 \label{Q_n}
 & \mathcal{Q}_n = \frac{1}{1 -  \mathcal{R}_n^2 \mathrm{e}^{2 i \varkappa_n L} }
\ .
\end{align}
\end{subequations}

Parameter $\mathcal{R}_n$ in Eq.~\eqref{R_n} obviously resembles the well-known expression for the reflection coefficient from the potential step, which is familiar from standard quantum mechanics textbooks (see, e.\,g., Ref.~\cite{bib:LandLifsh74}). It only differs from the latter coefficient by the slight modification (by substitution $\varkappa_n\rightarrow\beta_n$) related to periodicity of the potential on the one side of the step in the case under examination. In our particular problem, the stepwise jump of the effective potential results from the gradient renormalization of the mode energy while crossing the border between regions I or III of the waveguide and the corrugated region II (see Fig.~\ref{fig1}) wherein the mean value of the potential differs from zero.

\begin{figure}[h]
\begin{center}
  \scalebox{0.75}{\includegraphics{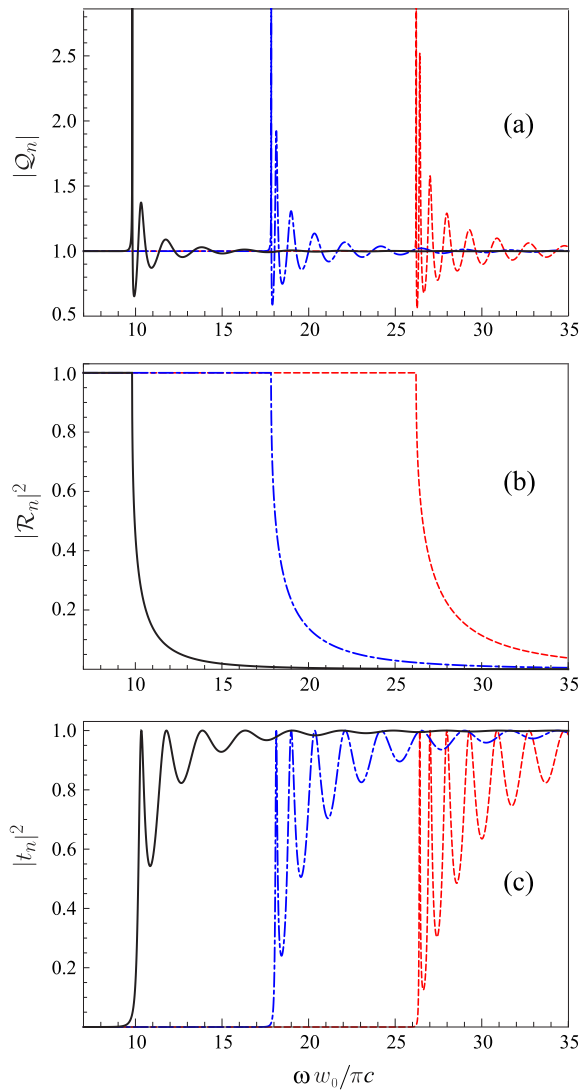}}
  \caption{(Color online) The plots of the quantities $|\mathcal{Q}_n|$, $|\mathcal{R}_n|^2$ and $|t_n|^2$ versus the frequency measured in units of $\pi c/w_0$. The points of $\mathcal{R}_n$ slumps in the upper plot correspond to the opening of the next waveguide channel. Geometrical parameters of the corrugated segment are chosen as follows: ${\Xi=3.33}$, $\sigma = 0.04\, w_0$, $L=0.3\, w_0$ .
\label{fig5}}
\end{center}
\end{figure}

The quantity $\mathcal{Q}_n$, which enters the Green function terms as a common multiplier, is of interest in itself, so it is worthwhile to examine it in more detail. This factor has a resonant character and is modulo maximal when the Bragg condition is met
\begin{equation}\label{Standing_waves}
  \varkappa_n L=l\pi\qquad (l\in\mathbb{N})\ ,
\end{equation}
which means that exactly half-integer number of the $n$'th mode wave lengths are accommodated on the length of the corrugated segment. In the plots of $|\mathcal{Q}_n|$ versus frequency (see Fig.~\ref{fig5}a), bearing in mind that through the Green function the mode density of states is expressed, one can see the well-known van~Hove singularities, which are related to the mode cutting-off in going across the boundary between straight and corrugated regions of the waveguide.

For the purposes of comparison, in Fig.~\ref{fig5} the dependencies on dimensionless frequency $\overline{\omega}= \omega(w_0/\pi c)$ are shown for three specific modes ($n=1,2,3$) of (a) resonance factor $\mathcal{Q}_n$, which determines the intensity of the field of the corresponding mode ``accumulated'' in the corrugated segment, (b) the modulo squared reflection coefficient from the boundary between corrugated and straight segments of the waveguide ($\mathcal{R}_n$), and (c) of the modulo squared coefficient $t_n$ of the mode transmission through the corrugated region. It is seen that at the points of appearance of the next propagating mode the mode density of states (proportional to $\mathcal{Q}_n$) has obvious singularity. To the left of the corresponding cutoff frequency ($\overline{\omega}_n^{(c)}$), the coefficient of mode reflection from the effective potential discontinuity ($\mathcal{R}_n$), in view of the mode assuming the evanescent character, is of the modulo one, whereas the transmission coefficient turns to zero. To the right of $\overline{\omega}_n^{(c)}$ the $\mathcal{R}_n$ modulus rapidly drops down, whereas the transmission factor abruptly rises, being at the same time subject to resonance oscillations \cite{Quasi-bound_states}. The ideal transparency of the waveguide in the oscillation maxima ($|t_n|=1$) corresponds to the appearance inside the corrugated segment of quasi-bound states, which arise due to the interference of waves \emph{partially} reflected from the interfaces between waveguide segments with a different mode structure. Thus, the corrugated segment of the waveguide plays actually the role of an open-ended resonator in which, due to its openness, the unlimited energy accumulation even at frequencies so close to the resonance ones is forbidden.

\subsection{The exact mode propagators}

The above obtained trial Green function may be thought of as a starting point for obtaining the exact mode propagator $G_{nn}(z,z')$ using the perturbation theory with regard to the intermode scattering. Such an interpretation of function $G^{(V)}_n (z,z')$ is corroborated by the fact that both of the Green functions are connected with one another by Lippmann-Schwinger equation, see Ref.~\cite{bib:Tar00}. The exact mode propagator may differ substantially from the trial one, since in Eq.~\eqref{GDIAG-FIN} the effective potential $\hat{\mathcal{T}}_{n}$ has appeared whose effect on the waveguide spectrum is not necessarily small. From the functional structure of this potential one can infer that within the framework of our theory the intermode scattering reduces to the trial mode intermixing, which is implemented by operator $\hat{\mathsf R}$ in Eq.~\eqref{T-oper}.

To analyze the intermode scattering in the arbitrary order of intermode potentials, with regard for the potential \eqref{T-oper} complex structure, is a rather difficult task. This can be accomplished efficiently only in the limiting cases of weak and strong trial mode intermixing. The intensity of the intermode scattering may be quantified by the operator $\hat{\mathsf R}$ norm, depending on whose value we may scrutinize both of the above indicated limits by expanding the $T$-potential in an appropriate manner.

\begin{figure}[h]
  \centering
  \scalebox{.7}{\includegraphics{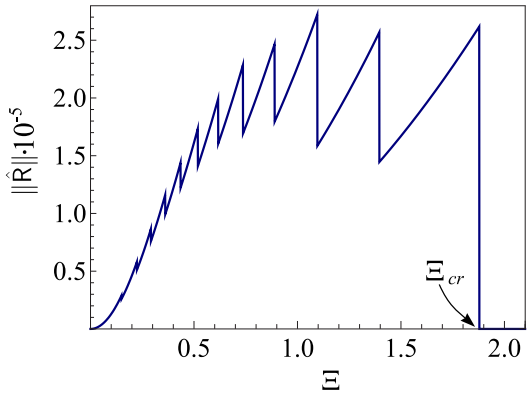}\qquad\qquad}
  \caption{The operator $\hat{\mathsf{R}}$ norm versus the degree of the corrugation sharpness. The comb-shaped structure of the graph is related to the sequential closing of the extended modes, in compliance with Eq.~\eqref{N_c0}. The waveguide parameters are as follows: $\sigma=0.05w_0$, $L=10w_0$, $N_{c0}=15$.
\label{fig6} }
\end{figure}

In Appendix~\ref{R_norm-estim}, the operator $\hat{\mathsf{R}}$ norm is estimated at an arbitrary value of the sharpness parameter $\Xi$. It is shown that with the chosen waveguide parameters, specifically, if the length of the corrugated segment exceeds considerably both the wavelength and corrugation period ($L\gg k^{-1},d$), while the parameter $\Xi$ does not exceed its critical value~$\Xi_{cr}$, the operator $\hat{\mathsf{R}}$ norm appears to have, for the most part, very large values, $\|\hat{\mathsf{R}}\|\gg 1$. It only assumes small values, as compared to unity, when either the corrugation is extremely smooth, so that $\Xi\ll(kL)^{-1}(d/L)$, or if parameter $\Xi$ has overcritical values, where the corrugated segment of the waveguide becomes the evanescent-mode one. However, in the dominant range of parameters, including the case where the corrugation is rather smooth ($\Xi\ll 1$),
the inequality holds $\|\hat{\mathsf R}\|\gg 1$, as is exemplified in Fig.~\ref{fig6} where the
numerical results are presented for the operator $\hat{\mathsf{R}}$ norm calculated for the
particular set of waveguide parameters.

At first glance, the fact that the trial modes become substantially intermixed, even if the corrugation is rather smooth, should imply that they cannot serve as a good approximation for determining the exact mode states. Yet, in the case of the large operator $\hat{\mathsf{R}}$ norm one can carry out the following formal manipulations with equation \eqref{GDIAG-FIN}. By considering this equation as the coordinate matrix element of operator equation
\begin{equation}\label{G_nn_oper_eq}
\left(\hat{\mathcal G}_n^{(V)-1} - \hat{\mathcal T}_n\right) \hat{G}_{nn} = \openone\ ,
\end{equation}
where $\hat{\mathcal G}_n^{(V)}$ is the operator whose coordinate matrix element is the Green function \eqref{TrialGreenFunction_2}, one can expand the inverse operator in Eq.~\eqref{T-oper} into a series not in the operator $\hat{\mathsf R}$ itself but rather in the its inverse, $\hat{\mathsf R}^{-1}$. The expression located between the projectors in Eq.~\eqref{T-oper} may be transformed in the following way,
\begin{align}\label{T-large_R}
  \hat{\mathcal U}
  \big(\hat{\openone}-\hat{\mathsf{R}}\big)^{-1} \hat{\mathsf{R}} &=
  \hat{\mathcal U} \Big[\hat{\mathsf{R}}^{-1} \big(\hat{\openone}-\hat{\mathsf{R}}\big)\Big]^{-1}=
  - \hat{\mathcal U} \big(\hat{\openone}-\hat{\mathsf{R}}^{-1}\big)^{-1} \notag\\&\approx
  - \hat{\mathcal U}  - \hat{\mathcal G}^{(V)-1} - \hat{\mathcal G}^{(V)-1} \hat{\mathcal U}^{-1}  \hat{\mathcal G}^{(V)-1}\ .
\end{align}
Taking into account that operator $\hat{\mathcal U}$ is of zero-diagonal nature in the mode variables, one can reduce the potential Eq.~\eqref{T-oper} to the form
\begin{equation}\label{T-strong_scatt}
  \hat{\mathcal T}_n \approx - \hat{\mathcal G}_n^{(V)-1} \Big[ \openone + \big( \hat{\mathsf{R}}^{-1}\big)_{nn}\Big]\ .
\end{equation}
By substituting it to equation \eqref{G_nn_oper_eq}, we arrive at the following operator expression for intramode propagator $G_{nn}(z,z')$:
\begin{equation}\label{Gnn-strong_scatt}
  \hat{G}_{nn} \approx \left[\openone + \frac{1}{2} \big( \hat{\mathsf{R}}^{-1}\big)_{nn} \right]^{-1} \frac{\hat{\mathcal G}_n^{(V)}}{2} \ ,
\end{equation}
or in the coordinate representation
\begin{equation}\label{Gnn-strong_sc}
  G_{nn}(z,z') \approx \frac{1}{2} {\mathcal G}_n^{(V)}(z,z') \ .
\end{equation}

Based upon this result it may be concluded that if the corrugation-induced intermode scattering becomes strong, in terms of the trial mode intermixing, the precise waveguide spectrum returns with parametric accuracy to the form specified by the initial trial Green function, which allows for the intramode scattering only. Considering the intermode potentials in Eq.~\eqref{G_mode-eqn} for $\|\hat{\mathsf{R}}\|\gg 1$ results in the twofold reduction of the amplitude of the wave excited by a given source in comparison to the amplitude of the wave we have chosen as the trial one.

Interestingly, although the intermode scattering in the system of our study is, chiefly, strong this should not have a strong impact upon the spectra of signals transmitted across the corrugated waveguide segment. As seen from relation \eqref{Gmn->Gnn}, at large operator $\hat{\mathsf{R}}$ norms for propagator $G_{mn}(z,z')$ with $m\neq n$ the operator (in $z$ coordinate) expression is valid
\begin{equation}\label{Gmn->R->Gnn}
  \hat{G}_{mn} \approx
  -\big(\hat{\mathsf{R}}^{-1}\big)_{mn}\hat{G}_{nn}\ .
\end{equation}
From Eq.~\eqref{Gmn->R->Gnn} it follows that if the waveguide is excited by the mode which is not of beyond-cutoff nature for corrugated segment, at the output of this segment the mode should be basically registered of the same mode index. The other modes should have small amplitudes at the exit of the corrugated region, in spite of the high-level intermode scattering within it. Although this conclusion seems to be unexpected at first sight, it is fully consistent with the data obtained, e.\,g., by Vellekoop and Mosk \cite{bib:VellekoopMosk08} in their studies of wave propagation through waveguides with a variable level of inhomogeneity, as well as with qualitative explanation of their results suggested later on in Ref.~\cite{bib:Pendry08}.

\section{Corrugation-induced rarefication of waveguide spectrum}

The above consideration suggests that regardless of the corrugation-induced intermode scattering being weak or strong, the spectrum of the corrugated waveguide is mainly determined by the set of \emph{trial} mode Green functions, i.\,e., it can be with good accuracy approximated by the entire  set of Bloch wave numbers,~$\varkappa_n$. For corrugated waveguide walls, the dependence of these wavenumbers on frequency $\omega$ bears the same qualitative character as for the ideally straight-wall waveguide. Yet, there is a number of essential differences between waveguides with straight and corrugated boundaries. One of the differences we have revealed is that the cutoff frequencies of the corrugated waveguide exceed the corresponding frequencies of the waveguide with straight walls.
\begin{figure}[h]
  \centering
  \scalebox{.7}{\includegraphics{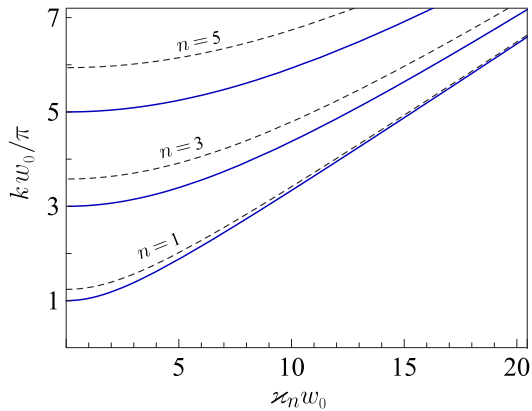}}
  \caption{(Color online) The Brillouin diagrams for corrugated (dashed lines) and the corresponding non-corrugated waveguide (solid lines). The corrugation parameters are: $\sigma=0.05\,w_0$, $\Xi=0.25$.
\label{fig7} }
\end{figure}
This can be seen from Fig.~\ref{fig7}, where Brillouin diagrams resulting from our formulas are presented for three different modes of the corrugated and the corresponding non-corrugated waveguide.

Besides, even small-amplitude corrugation of the waveguide walls has a noticeable effect upon the density of its cut-off frequency spectrum. Without corrugation, the set of these frequencies is equidistant in mode numbers, $\omega_n^{(c)}=\pi nc/w_0$. If corrugation is present, the equidistance is violated and the interval between the neighboring cut-off frequencies increases with a growth in the corrugation sharpness. This is
\begin{figure}[h!]
  \vspace{12pt}
  \centering
  \scalebox{.7}{\includegraphics{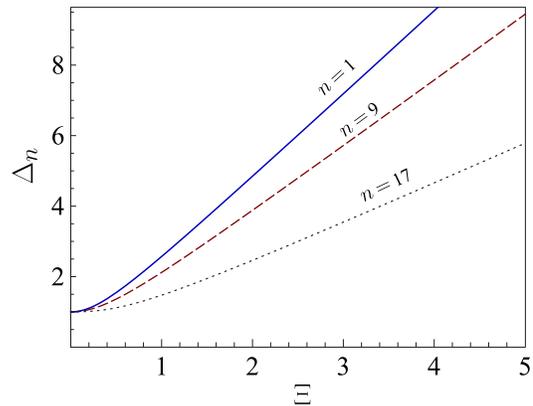}}
  \caption{(Color online) The relative interval between neighboring cut-off frequencies for the corrugated waveguide against the corrugation sharpness. Parameter $\sigma$ is fixed and equal to $0.05\,w_0$.
\label{fig8} }
\end{figure}
clearly evident from Fig.~\ref{fig8}, where the relative interval between neighboring cut-off frequencies is depicted, namely,
\begin{equation}\label{Cut-off_rel}
  \Delta_n=\frac{\Omega_{n+1}^{(c)}-\Omega_n^{(c)}}{\omega_{n+1}^{(c)}-\omega_n^{(c)}}\ ,
\end{equation}
where $\Omega_n^{(c)}$ is the $n$-th mode cut-off frequency for the corrugated waveguide. One can see from the graphs that the spectrum of modes propagating in the waveguide with corrugated walls is considerably rarefied with an increase in the corrugation sharpness, as compared to the similar spectrum of the waveguide of the same average width whose side boundaries are ideally smooth.

The result we have obtained is in conformity with the general claim regarding the normal frequencies of conservative two-dimensional systems \cite{bib:Rabinivich84}, namely, that the addition of couplings into a dynamical system can only enlarge the intervals between its normal frequencies. In our particular case, this statement turned out to be valid for infinite-dimensional dynamical system \eqref{G_mode-eqn}, where the role of coupling is played by the level of corrugation. Note that the effect of the cut-off frequency shift to a higher-frequency region was already noted in Ref.~\cite{bib:Esteban91}. Yet, it was not appropriately examined therein and no general rules were established.

One more intriguing property of the corrugated waveguide is closely related to the cut-off frequency growth, namely, a decrease in the number of modes that are allowed to pass through the waveguide, as the sharpness of corrugation increases with its amplitude being kept constant. As can be seen from Eq.~\eqref{N_c0}, a decrease in the number of extended modes with increasing the sharpness parameter $\Xi$ may be quite significant, up to the waveguide being entirely ``locked out'', while the number of modes being allowed to propagate in the non-corrugated regions is reasonably large. This property may be used for creating the waveguide filters, as it permits quite significant spectral modifications of signals transmitted through artificially corrugated waveguides.

\begin{widetext}
%
\section{The energy transport in the waveguide with corrugated segment}

The Green function obtained in Section \ref{Green_func_calc}, apart from the spectral properties of the waveguide with corrugated insert, allows for direct calculations of its conducting properties, in particular, of power transmission. The power radiated by the plane antenna positioned at cross-section $S_{z'}$ and then transferred through cross-section $S_z$ at the point with longitudinal coordinate $z$, being averaged over oscillation period is given by formula
\begin{align}\label{Av_Power-1}
  \overline{P(z,z')} =-\,\bm{\mathcal{C}}k\,\,\mathrm{Im}\!\int\limits_{S_z}ds\int\limits_{S_{z'}}ds'
  G(\mathbf{r},\mathbf{r}')\frac{\partial}{\partial z}G^*(\mathbf{r},\mathbf{r}')
  = -\,\big(\bm{\mathcal{C}}l_x^2\big)k\,\,\mathrm{Im}\!\!\!\!\!\int\limits_{-w(z)/2}^{w(z)/2}\!\!\!dy\!\!\!
  \int\limits_{-w(z)/2}^{w(z)/2}\!\!\!dy'
  G(\mathbf{r},\mathbf{r}')\frac{\partial}{\partial z}G^*(\mathbf{r},\mathbf{r}')\ .
\end{align}
Here, $\bm{\mathcal{C}}$ is the dimensional constant, whose presence relates to the fact that the dimensionality of Green function from equation \eqref{Helmholtz_Green} does not meet that of the electric field from Maxwell's equations; $l_x$ is the waveguide dimension along axis $x$, the asterisk denotes complex conjugation. By substituting Green function in the form of double series \eqref{Green_ModeRep} into Eq.~\eqref{Av_Power-1} we arrive at the expression
\begin{align}\label{Av_Power-2}
  \overline{P(z,z')}= & -\big(\bm{\mathcal{C}}l_x^2\big) k\,\mathrm{Im}\sum_{m,n=1}^{\infty}
  G_{mn}(z,z')\frac{\partial}{\partial z}G_{mn}^*(z,z')\notag\\
  & -\big(\bm{\mathcal{C}}l_x^2\big)\frac{w'(z)}{w(z)}k\,\mathrm{Im}\sum_{n=1}^{\infty}
  \sum_{\genfrac{}{}{0pt}{1}{i,k=1}{(i\neq k)}}^{\infty}
   G_{in}(z,z') G_{kn}^*(z,z')\frac{2ik}{i^2-k^2}\left[(-1)^{i+k}+1\right]\ .
\end{align}

In the context of our problem the second term in the r.h.s. of Eq.~\eqref{Av_Power-2} may be omitted. This is substantiated, first, by the observation that this term does not contain the products of the diagonal mode Green functions. As far as the off-diagonal ones are concerned, we neglect them in view of equality \eqref{Gmn->R->Gnn}. The additional argument to neglect the second term in the r.h.s. of Eq.~\eqref{Av_Power-2} is the presence of factor $w'(z)/w(z)$, which is exactly zero if the receiver is positioned outside the corrugated region. With allowance for the result \eqref{Gnn-strong_sc} we thus arrive at the following formula for the power transmitted across the corrugated segment of the waveguide,
\begin{align}\label{Av_Power-main}
  \overline{P}_L(\Xi,\omega)=-\big(\bm{\mathcal{C}}l_x^2\big) \frac{\omega}{4c}\,\mathrm{Im}\sum_{n=1}^{\infty}
  {\mathcal G}_n^{(V)}\big(z,z'\big)\frac{\partial}{\partial z}{{\mathcal G}_n^{(V)}}^*\big(z,z'\big)\Bigg|_{\genfrac{}{}{0pt}{}{z=L/2\phantom{2}}{z'=-L/2}}=
  \big(\bm{\mathcal{C}}l_x^2\big) \frac{\omega}{4 c}\,\ \sum_{n=1}^{N_c(\Xi)} \frac{|t_n|^2}{4 k_n}\ .
\end{align}

To make a detailed calculations of this power, one should substitute either expression \eqref{TrialGreenFunction_2} for the trial Green function or the transmission coefficient Eq.~\eqref{t_n} into formula \eqref{Av_Power-main}.
In figs.~\ref{fig9} and \ref{fig10}, the power transferred through the corrugated segment of the waveguide is depicted as a function of the normalized frequency and the sharpness parameter,~$\Xi$. The stepwise form of both of the graphs is related to the addition (or, correspondingly, cutting-off) the effective (gradient-renormalized) extended modes in the waveguide spectrum, in 
%
%
\begin{figure}[h!]
  \begin{minipage}[b]{.45\linewidth}
  \scalebox{1}{\includegraphics[height=5.5cm]{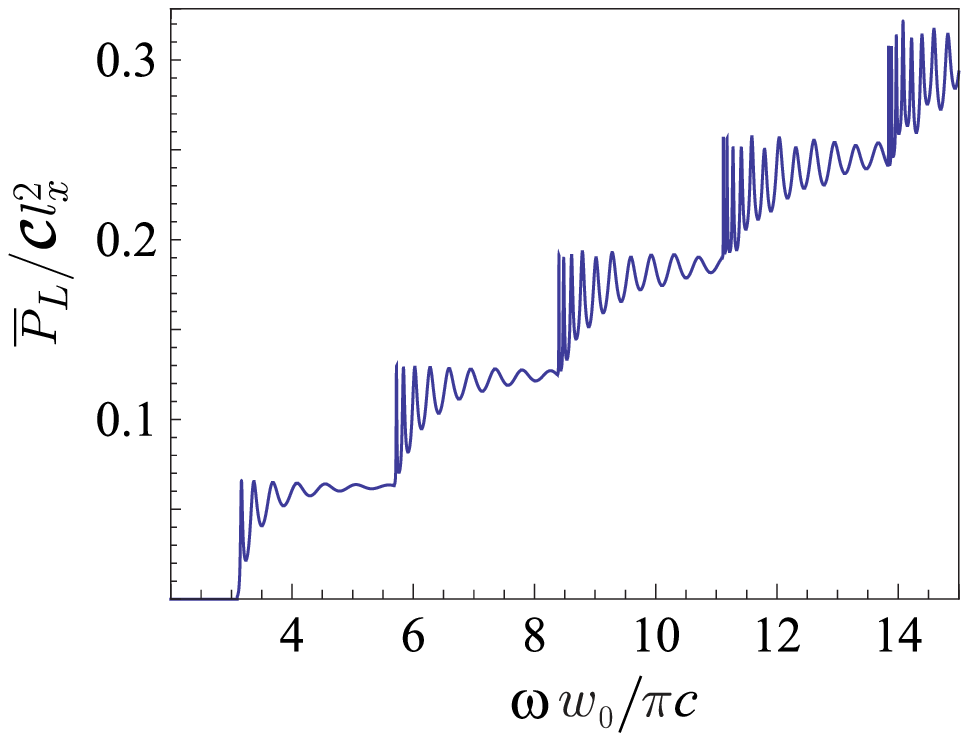}\qquad}
  \caption{(Color online) Frequency dependence of the power transmitted across the corrugated waveguide segment ($\sigma= 0.02w_0$, $L=1.5w_0$, $\Xi =1$).
\label{fig9} }
 \end{minipage}
\hspace{1cm}
 \begin{minipage}[b]{.45\linewidth}
  \scalebox{1}{\includegraphics[height=5.5cm]{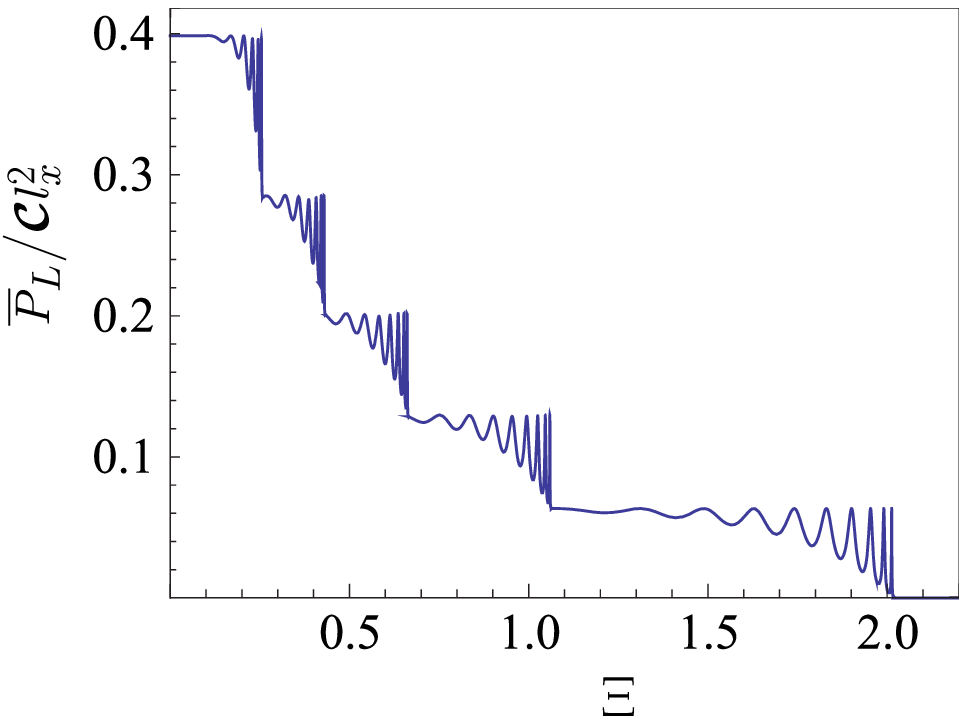}\qquad}
  \caption{(Color online) The transmitted power against the corrugation sharpness ($\sigma=0.02w_0$, $L=2w_0$, $\omega=6\pi c/w_0$).
\label{fig10} }
 \end{minipage}
\end{figure}
%
%
%
accordance with formula~\eqref{N_c0}. However, whereas the steps in Fig.~\ref{fig9} are the well-known physical fact, the analogous stepwise behavior of the transmitted power as a function of corrugation sharpness, shown in Fig.~\ref{fig10}, is predicted therein for the first time. The oscillations seen on the plateaus in both of the graphs, \ref{fig9} and \ref{fig10}, are related to Bragg resonances in the transmission of the marginal waveguide mode.
\end{widetext}
%

\section{Summary and discussion}

To conclude, we formulate the main results obtained in the present paper and give some comments on the exploited calculation technique. We have developed the theory of wave transmission through the waveguide containing the finite segment with periodically corrugated side boundaries. Such an inhomogeneous insert in the initially unfigured waveguide can have a considerable impact on its spectral properties. The central role in the modification of the entire waveguide mode structure is played by the sharpness of the insert corrugation. Even though the boundary asperities might be of very small amplitude, the number of extended modes in the waveguide is substantially reduced when the corrugation is sharpened, reaching zero at some critical value of the governing sharpness parameter.

To solve the problem we have applied the method of oscillation mode separation, which was developed earlier with regard to waveguide-like systems with bulk inhomogeneities. In the present study we have adjusted this method to waveguides with inhomogeneous boundaries by using thereto the local transverse-mode representation, which makes it possible to present surface inhomogeneity as the bulk one.

The separation of waveguide modes, which is basically equivalent to the separation of variables in the original wave equation, was carried out through the specific operator procedure essentially equivalent to the construction of the perturbation series with regard to the intensity of the intermode scattering. In developing this theory, we started from the solution to the auxiliary problem, in which the intermode scattering is formally disregarded while the intramode scattering is taken into account in full. The application of this procedure has enabled us to find that the intermode scattering induced by the corrugated waveguide segment is normally very strong, resulting actually in the mode system homogenization. In these conditions, the spectrum of sharply uneven waveguide is restored to the state were the intermode scattering is entirely absent while the intramode scattering is accounted for. The actually large strength of the intermode scattering reveals itself in the twofold reduction of the field amplitude in comparison with the case were this scattering is disregarded arbitrarily.

By solving the set of decoupled equations derived using our method, we succeeded in showing that the corrugated segment of the waveguide may be considered (by analogy with quantum particle transport) as the modulated potential barrier, whose width is coincident with the actual length of the segment whereas the height is prescribed by the level of corrugation sharpness. Inside this sharpness-controlled barrier the field is transferred in a resonant manner by means of the quasi-steady states arising due to the multiple wave reflection from the barrier ends. The ``external'' waveguide modes falling upon the barrier are mutually re-scattered while passing through it, but at the exit point their arrangement in mode indices is basically restored. As the barrier increases with growing the corrugation sharpness, the propagating modes are successively cut off, though the asperity height is insufficient for this purpose. The mode cutting-off leads to rarefication of the entire (combined) waveguide spectrum, so that at some critical level of the asperity sharpness the waveguide becomes the evanescent-mode one.

Another important result of the application of our method is that by using it one can analytically trace the overall dynamics of the propagation constants against the sharpness of boundary corrugation. Although functional dependence of the longitudinal wavenumber on the frequency for both the straight and the corrugated waveguide remains the same, in the presence of corrugation the cut-off frequencies line up non-equidistantly. The shift of these frequencies to the higher-frequency range grows considerably both with a decrease in the corrugation period and with a mode number increase. The difference between neighboring cut-off frequencies increases as well, so that the waveguide spectrum becomes significantly more sparse.

The non-equidistance of the cut-off frequency set may be utilized in practice to achieve the single-frequency mode in microwave amplifiers, in developing the waveguide-based filters for complex signals, as well as for narrow-band generators with Bragg reflectors. In such generators, the amplification curve normally has finite width. Therefore, in order to prevent the given mode interaction with higher modes it is necessary that they do not fall into the amplification band. At the same time, the operating mode frequency, at which the output-input ratio reaches a maximal value, is situated at a relative distance of approximately 1/4 from the amplification curve lower bound. So, it is quite desirable for the cut-off frequencies to be shifted disproportionately to mode numbers.

\appendix

\section{Estimation of the mode-mixing operator norm}
\label{R_norm-estim}

According to the standard definition of the operator norm \cite{bib:KolmFom68,bib:Kato66}, we have to calculate the quantity
\begin{equation}\label{Norm-def}
  \|\hat{\mathsf{R}}\|^2=\sup_{\psi}
  \frac{\big(\hat{\mathsf{R}}\psi,\hat{\mathsf{R}}\psi\big)}{\big(\psi,\psi\big)}\ ,
\end{equation}
where trial functions $\psi(z)$ must belong to the functional space where operator $\hat{\mathsf R}$ is active. The domain of definition of this operator, as shown in Sec.~\ref{ModeSep}, is the coordinate-mode space, so the result of its action on some scalar function (in particular, on the single component of a vector function) is a vector.

The support of potentials \eqref{Potentials_VnUnm} is the segment $\mathbb{C}: |z|<L/2$. Therefore, it is exactly this particular segment that we will consider as the support for the entire set of trial functions in Eq.~\eqref{Norm-def}. The finite-support functions are Fourier-expandable, so in the capacity of trial functions in the definition Eq.~\eqref{Norm-def} it is reasonable to use the set of exponential functions $\psi(z)=\exp(iqz)$. The compactness of the support for the latter functions may be ensured by choosing the integration contour for Fourier integrals to be not the exacly real axis, $q\in\mathbb{R}$, but rather the horizontal line in the plane of the complex variable ($q$) properly offset to the upper or lower half-plane, e.\,g., $q\in\mathbb{R}\pm i/L$.

It is convenient to proceed with the norm calculation in the selected class of functions not for the operator itself ${\hat{\mathsf{R}}=\hat{\mathcal{G}}^{(V)}\hat{\mathcal{U}}}$ but rather for its transposed counterpart, where operators $\hat{\mathcal{G}}^{(V)}$ and $\hat{\mathcal{U}}$ are rearranged. Considering that the norm squared of the chosen trial function is of the order of $L$, we actually need to estimate (by an order of magnitude) the expression
\begin{align}\label{Norm-new_trial}
\displaybreak
  \|\hat{\mathsf{R}}\|^2= & \sup_{q\in\mathbb{R}+i/L}
  \frac{1}{L}\int_{\mathbb{C}}dz\sum_{i,k\neq n}\iint_{\mathbb{C}}dz_1dz_2
  \hat{U}_{in}^*(z)\hat{U}_{kn}(z)\notag\\
 & \times {G_{i}^{(V)}}^*(z,z_1)G_{j}^{(V)}(z,z_2)\mathrm{e}^{-iq(z_1-z_2)}\ .
\end{align}

The potential $\Delta V_n(z)$, see Eq.~\eqref{V_n-oscillate}, does not modulo exceed the average potential $\overline{V_n(z)}$. Hence, we can simplify the integrand in Eq.~\eqref{Norm-new_trial} by setting all the factors $g(z|n)\equiv 1$ in expression \eqref{TrialGreenFunction_2} and replacing the quantity $\varkappa_n$ in the exponentials by the wavenumber ${\widetilde{k}_n=\sqrt{k_n^2 - \overline{V_n(z)}}}$. The reflection coefficient \eqref{R_n} is not modulo larger than unity (see Fig.~\ref{fig5}), so instead of exact expression \eqref{TrialGreenFunction_2} we will use in what follows its truncated version, namely,
\begin{equation}\label{G(V)-model}
  G_{n}^{(V)}(z,z')\approx\frac{-i\mathcal{Q}_n}{2\widetilde{k}_n}
  \exp\left(i\widetilde{k}_n|z-z'|\right)  \ .
\end{equation}
With this expression, only two first terms in Eq.~\eqref{TrialGreenFunction_2} are actually allowed for, whose amplitude factors are $\mathcal{G}_{11}^{(n)}$ and $\mathcal{G}_{22}^{(n)}$. In addition, in the right-hand sides of Eqs.~\eqref{G_11} and \eqref{G_22} we retain the terms without the factors of $\mathcal{R}_n$, bearing in mind that the order-of-magnitude estimation only is of importance. All the unrecorded terms contribute to the norm \eqref{Norm-def} additively with function \eqref{G(V)-model}, so they cannot affect significantly the estimates obtained below.

As far as the intermode potentials in Eq.~\eqref{Norm-new_trial} are concerned, their expression \eqref{Unm} contains two functionally different terms (with zero and non-zero average value). We will consider their contribution to the mode-mixing operator norm separately. The contribution of the first, quadratic in $w'(z)$ term, after the function \eqref{G(V)-model} being represented in terms of the Fourier integral, is written as
\begin{widetext}
\begin{align}\label{Norm_first}
   \|\hat{\mathsf{R}}\|_1^2=\sup_{q\in\mathbb{R}+i/L}
   \frac{1}{L}\left(\frac{\Xi}{w_0}\right)^4\big|\mathcal{Q}_n\big|^2
   \sum_{i,k\neq n} & (B_{in}D_{in})(B_{kn}D_{kn})\notag\\
  \times & \iint\limits_{-\infty}^{\infty}\frac{ds_1ds_2}{(2\pi)^2}  \int_{\mathbb{C}}dz
   \frac{\mathrm{e}^{-i(s_1-s_2)z}}{(s_1^2-\widetilde{k}_n^{2*})(s_2^2-\widetilde{k}_n^{2})} \iint_{\mathbb{C}}dz_1dz_2\mathrm{e}^{-i(q-s_1)z_1}\mathrm{e}^{i(q-s_2)z_2}\ .
\end{align}
\end{widetext}
Here, the integral over $z$ is calculated without difficulty if the length $L$ is sufficiently large. Specifically, when $|\widetilde{k}_n|L\gg 1$ this integral may be thought of as the prelimit $\delta$-function,
\begin{equation}\label{Delta-prelimit}
  \Delta(\kappa)=\int_{\mathbb{C}}dz\exp(\pm i\kappa z)=
  \frac{\sin(\kappa L/2)}{\kappa/2}\xrightarrow[L\to\infty]{}2\pi\delta(\kappa)\ .
\end{equation}
Computing then the remaining integrals in Eq.~\eqref{Norm_first} and taking into account that the major contribution to the sums over $i$ and $k$ comes from the terms with the extended-type trial Green functions, we arrive at the estimate
\begin{equation}\label{||R||_1-prefin}
  \|\hat{\mathsf{R}}\|_1^2\sim\Xi^4\left(\frac{L}{w_0}\right)^4 N_c^4(\Xi)\sim
  \left(\frac{\Xi^2}{1+\Xi^2}\right)^2(kL)^4\ .
\end{equation}

The contribution to norm \eqref{Norm-new_trial} of the second, linear in $w'(z)$ term of potential \eqref{Unm} is calculated in a similar manner, and the estimate of this contribution reads
\begin{equation}\label{||R||_2-prefin}
  \|\hat{\mathsf{R}}\|_2^2\sim\Xi^2\left(\frac{L}{w_0}\right)^4
  \left(\frac{w_0}{d}\right)^2N_c^2(\Xi)\sim
  \frac{\Xi^2}{1+\Xi^2}(kL)^2\left(\frac{L}{d}\right)^2\ .
\end{equation}
By comparing the right-hand sides of Eqs.~\eqref{||R||_2-prefin} and \eqref{||R||_1-prefin} one can reveal that their ratio equals to the value of $(1+\Xi^2)/(\Xi kd)^2$. Under the conditions suggested by inequalities Eq.~\eqref{sigma<<w_0/Nc}, this parameter is greater than unity, so the result \eqref{||R||_2-prefin} may be reckoned as the final estimate of the operator $\hat{\mathsf R}$ norm in the slope parameter region $\Xi<\Xi_{cr}$, where the intermode scattering is ensured by extended modes. In the region $\Xi>\Xi_{cr}$ all the trial modes become evanescent, so it does not make any sense to take account of their mutual intermode scattering.




\begin{thebibliography}{cc}
\bibitem{bib:refto_Ignatovich86}
Expressions \eqref{t_n} and \eqref{r_n} are coincident with transmission and reflection factors obtained by Ignatovich Ref.~\cite{bib:Ignatovich86}, who studied the mathematical aspects of wave propagation in periodic potentials with finite number of periods.
\bibitem{Quasi-bound_states}
The analogous transmittance oscillations are well known in the theory of wave propagation through finite periodic structures (see, e.\,g., Ref.~\cite{bib:Londergan99}). Yet, in our case the appearance of these oscillations is not directly related to the corrugation periodicity and to the formation of different Bloch spectral bands. In our theory, the resonances in the transparency arise when new effective (gradient-renormalized) extended mode is opened in the corrugated segment and, depending on the length of the segment, the conditions for the formation of the quasi-bound states of Bragg nature inside it are realized. The density of the transmission resonances on the frequency axis is growing rapidly with increase in the length of the corrugated segment. Thus, in the experiments this may be taken merely as the transmission curve ``spreading'' near each of the cut-off frequencies.
\bibitem{Dephasing}
The dephasing effect produced by the intermode scattering was fist noticed in Ref.~\cite{bib:Tar00}.
\bibitem{On_Chaos}
According to Chirikov's criterium \cite{bib:Chirikov79}, the chaoticity level is related to the overlap of resonances, which evidently grows with increasing the resonance line width.
\end{thebibliography}
\end{document}